\documentclass[twocolumn,aps,prl,tightenlines,floats]{revtex4}
\def\bea{\begin{eqnarray}}
\def\eea{\end{eqnarray}}
\def\etal{{\it et al.\/}}
\usepackage{graphics}
\usepackage{graphicx}
\usepackage{epsf} 
\usepackage{amsmath}
\usepackage{amssymb}
\usepackage{bbold}
\begin{document} 

%\begin{widetext}

%\preprint{
\phantom{0}
\vspace{-1.4in}
\hspace{6in}\parbox{1.5in}{ \leftline{JLAB-THY-04-39}
                \leftline{WM-04-119}
			             %\leftline{nucl-th/04?????}
               % \leftline{}\leftline{}\leftline{}\leftline{}
\vspace{-3.6in}}  % moves the preprint box down
%}
\title
{\bf The shape of the nucleon}

\author{Franz Gross$^{1,2}$ and Peter Agbakpe$^{3}$
\vspace{-0.1in}  }

\affiliation{
%\address{
$^1$College of William and Mary, Williamsburg, Virginia 23185 \vspace{-0.15in}}
\affiliation{
$^2$Thomas Jefferson National Accelerator Facility, Newport News, VA 23606 \vspace{-0.15in}}
\affiliation{
$^3$Norfolk State University, Norfolk, VA 23504
}

%\begin{widetext}
%\end{widetext} 

\date{\today}

\begin{abstract} 

We show that all four of the nucleon form factors can be very well explained using the manifestly covariant spectator theory.  The nucleon is modeled as a spherical state of three constituent quarks with electromagnetic form factors, all determined by the fit to the data.  
%\maketitle

\end{abstract}
 
\phantom{0}
%\vspace{7.0in}
%\vspace{-6in}
\vspace*{0.9in}  % sets how far the title is below the preprint box

\maketitle

%%%%%%%%%%%%%%%%%%%%%%%%%%%%%%%%%%%%%%

%\section{Introduction} 

There have been many theoretical attempts to explain the recent high precision polarization transfer measurements of the ratio of proton form factors $G_{Ep}/G_{Mp}$ \cite{Jones,Gayou} and neutron form factors $G_{En}/G_{Mn}$ \cite{Zhu,Warren,Madey}.  As shown in Ref.~\cite{Madey}, recent calculations that model the microscopic structure of the nucleon using the Skyrme model \cite{Holzwarth:1996xq}, front-form quantum mechanics \cite{Cardarelli:2000tk,Ma:2002ir,Miller:2002ig,Miller:2003sa},  or point-form quantum mechanics \cite{Boffi:2001zb} fail to fit {\it all\/} of the form factor data.  Models inspired by vector dominance with 7 to 12 parameters do much better \cite{Lomon:2002jx,Bijker:2004yu}, but do not model the quark structure of the nucleon.  These difficulties  have lead some to suggest  \cite{Miller:2003sa,Buchmann:2001gj} that the proton is deformed, and this possibility has been widely discussed, reaching Science News \cite{SN94} and USA Today \cite{USA}.  While it is possible for a spin 1/2 system to be deformed \cite{Buchmann:2001gj}, the accompanying rotational bands and large spectroscopic factors have not been observed.

In this letter we use a simple model to show that the beautiful Jefferson Lab data do not require that the proton be anything other than spherical.  Our model assumes that the nucleon is composed of three valence {\it constituent\/} quarks (CQ), and that the main role of the surrounding sea of gluons and quark-antiquark pairs is to dress these CQ.  Through this dressing the CQ acquire mass, size, and structure, and are composite systems with electromagnetic form factors of their own.  The structure of these form factors is given by QCD, which requires that the quarks become point particles as the momentum transfer $Q^2\to\infty$, and that the $\gamma^*\to q\bar q$ spectral function at finite $Q^2$ be the sum of the $q\bar q$ bound state poles and resonances allowed by confinement (vector dominance).  

We use the covariant spectator formalism \cite{Gross:1969rv} to  model the bound states of three CQ.  Using this method, {\it all\/} of the Poincar\'e transformations (including rotations and boosts) can be handled exactly,  important when discussing the shapes of states.  This formalism has already been used successfully to calculate the  relativistic wave function of $^3$H  \cite{Gr82,SG1,SG2}, which is described by a vertex function with two of the three hadrons on mass-shell, and the third off-shell.  The wave function has the form
\bea
\Psi\equiv\left<p_2,p_3|\psi|P\right>=\frac{1}{m_q-\not\!p_1}\left<p_2,p_3|\Gamma |P\right> \label{eqa}
\eea
with $m_q$ the mass of the (dressed) quark, $P=p_1 + p_2+p_3$, $\Gamma$ the vertex function, and $p_2^2=p_3^2=m_q^2$.  We adopt a simple approximation in which the distributed mass of the spectator diquark system, $p^2\equiv(p_2+p_3)^2$, is fixed at a mean value of $m^2_s$.  Both $m_s$ and $m_q$ disappear from the final result.  

If the mass of the nucleon is larger than the mass of the three quarks, $M>3m_q$, the pole in the propagator of the off-shell particle 1 moves to the real $p_{10}$ axis.  If this pole were not cancelled by a zero in the vertex function, it would describe free scattering of the quarks.  However, it has been shown that a dynamical consequence of confinement is to automatically produce a zero in the vertex function that  cancels the pole \cite{Savkli:1999me}.  In this work we include confinement by modeling the wave function directly, assuming that (\ref{eqa}) is smooth at the point $p_1^2=m_q^2$.  
 
In the symmetric quark model the wave function can be written as the sum of two components
\bea
\Psi_{\rm NR}=\left(\cos\theta\;\phi_I^0 \phi_s^0 +\sin\theta \;\phi_I^1\phi_s^1\right)\psi\, ,\label{eq1}
\eea
where $\psi$ is a scalar function symmetric in the coordinates, and $I=\pm\frac{1}{2}$, $s=\pm\frac{1}{2}$ are isospin and spin projections, respectively.  The states $\phi^J_I$ couple quarks 2 and 3  to total isospin $J=0$ or 1:
\bea
\phi_I^0=\xi_{_{23}}^{0}\,  \chi^I_{_1}\, ,  \qquad\quad
\phi_I^1=-\frac{1}{\sqrt{3}}
\left(\tau\cdot\xi^{1}_{_{23}}\right)\chi^I_{_1} \, , 
\label{eq2}
\eea
with $\chi_1^{\frac{1}{2}}=u_1$, $\chi_1^{-\frac{1}{2}}=d_1$,  $\xi_{_{23}}^0= \frac{1}{\sqrt{2}}\left(u_2d_3-d_2u_3\right)$, and $\xi^{1+}_{_{23}}= u_2u_3$,  $\xi^{10}_{_{23}}= 
 \frac{1}{\sqrt{2}}\left(u_2d_3+d_2u_3\right)$, and 
$\xi^{1-}_{_{23}}=d_2d_3$.
The spin wave functions have the same form.   The sum (\ref{eq1}) is symmetric under the interchange of quarks 2 and 3, and when $\cos\theta=\sin\theta$, Eq.~(\ref{eq1}) is also symmetric under interchange of all the quarks, so in the following we discard the quark labels 1, 2, and 3.   

Following (\ref{eq1}), we assume that the {\it relativistic\/} wave function (\ref{eqa})  has the {\it manifestly covariant\/} form 
\bea
\Psi_{_\alpha} &=&\frac{1}{\sqrt{2}} u_{_{\alpha}}(P,s)\, \xi_{_{}}^{0}\,\chi^I \;\psi_0(P,p) \cr
&+&\frac{1}{3\sqrt{2}} (\gamma_5 \gamma_\mu)_{_{\alpha\beta}}\,\eta^{\mu} u_{_{\beta}}(P,s) \left(\tau\cdot\xi^{1}_{_{}}\right)\chi^I_{}\psi_1(P,p) \qquad
\label{eq4}
\eea
where $p$ is the on-shell diquark four-momentum ($p^2=m_s^2$), $u(P,s)$ is the spinor of the nucleon, $\alpha$ is the Dirac index of the off-shell quark, and $\eta$ is the covariant polarization vector of the spin-one diquark.   The scalar functions $\psi_0$ and $\psi_1$ can depend on only one variable, which we choose to be
\bea
\chi=\frac{M^2+m_s^2 -(P-p)^2}{Mm_s}\, .
\eea
In the cm frame, $\chi=2E_s(\kappa)$, where $E_s(\kappa)=\sqrt{1+\kappa^2}$ and $\kappa$ are the energy and magnitude of the three-momentum ${\bf p}$ of the on-shell diquark in units of $m_s$.  Note that the wave function depends only on the magnitude of ${\bf p}$, and hence is spherically symmetric.  To get the correct asymptotic $Q^{-4}$ dependence of the nucleon form factors predicted by QCD, we choose a simple double pole dependence for the scalar function $\psi_0$
\bea
\psi_0(P,p)=\frac{N_0}{m_s(\beta_1-2 +\chi)(\beta_2-2 +\chi)}
\quad
\eea
where $\beta_1$ and $\beta_2$ are range parameters in units of $Mm_s$ and $N_0$ is a normalization constant.   In the  nonrelativistic (NR) limit ($\kappa<\!\!<1$), $\psi_0$ becomes the familiar Hulthen wave function
\bea
\psi_0 \xrightarrow[{\rm NR}] {}\; \frac{N_0}{m_s
(\beta_1 + \kappa^2)(\beta_2 + \kappa^2)}\, .
\eea
We chose the form of $\psi_1$ to be related to $\psi_0$ by a relativistic factor that approaches unity in the nonrelativistic limit
\bea
\frac{\psi_1(P,p)}{\psi_0(P,p)}=\frac{1}{\sqrt{\frac{1}{3}+\frac{\chi^2}{6}}} \; \xrightarrow[{\rm NR}] {}\;  1\, .  \label{BA}
\eea
This choice insures that both the proton and the neutron charges are correctly given by a single normalization condition (see Eq.~(\ref{norm}) below), and is a relativistic generalization of the symmetry requirement $\psi_1=\psi_0$. With these definitions, the NR limit of $\Psi_\alpha$  in its cm frame becomes $\Psi_{\rm NR}$.

With these ans\"atze, the form factors of a nucleon with isospin $I$ are given in the spectator theory by the relativistic impulse approximation (RIA) \cite{Gross:2003qi}
%\begin{widetext}
%
\bea
J^\mu_I %=
&=&
F_{1}(Q^2)+ F_{2}(Q^2)\frac{i\sigma^{\mu\nu}q_\nu}{2M} \cr&&\cr
&=&
%=
3\,\bar u(P_+,\lambda') {\cal J}_{I}^\mu \,u(P_-,\lambda)\, ,
\eea
where $F_{i}$ are the nucleon Dirac and Pauli  form factors and   ${\cal J}^\mu_{I}$ is the current of a single quark  with isospin $I$
\bea
{\cal J}_{I}^\mu =\frac{1}{2}\int \frac{m_s^3\,d^3
\mathbb{\kappa}}{(2\pi)^3 m_s 2E_s(\kappa)} \Bigg\{j_{I}^\mu \psi_0(P_+,p)\psi_0(P_-,p)\qquad\cr
 -\frac{1}{9} \gamma_\nu \gamma_5 \tau_j j_{I}^\mu
\tau_j \gamma_5\gamma_{\nu'} \Delta^{\nu\nu'}\!
\psi_1(P_+,p)\psi_1(P_-,p)
\Bigg\} ,\quad \label{eq8}
\eea
%
%\end{widetext}
and we define $P_\pm=P\pm \frac{1}{2}q$.   The CQ current, $j_{I}^\mu$, is defined in Eq.~(\ref{CQc}) below.   We have summed over the polarization vectors of the diquark, using
\bea
\sum_\eta \eta^{\nu*} \eta^{\nu'}\equiv \Delta^{\nu\nu'} = 
-g^{\nu\nu'} + \frac{p^\nu p^{\nu'}}{m_s^2}\, .
\eea
The total result is three times the contribution of one quark.

The manifestly covariant current ${\cal J}^\mu_I$ is most easily calculated in the Breit frame with $Q_0= Q/M$, $P_0=\sqrt{1+Q_0^2/4}$ and $P_\pm=M\{P_0, 0,0, \pm Q_0/2\}$.  In this frame $\psi_0$ becomes
\bea
\psi^\pm_{0}\equiv\psi_0(P_\pm,p)
=\frac{N_0}{m_s \,D_\mp(Q)} \label{eq9}
\quad
\eea
with $D_\pm(Q)=(\beta_1-2+\chi_\pm)(\beta_2-2+\chi_\pm)$ and 
$\chi_\pm=2 P_0E_s(\kappa)\pm \kappa \, Q_0\, z$, 
where $z$ is the cosine of the angle between ${\bf p}$ and ${\bf q}$.  Similarly, the $\psi^\pm_1$ wave functions are obtained from Eq.~(\ref{BA}) by substituting $\chi\to\chi_\mp$.

The most general form of the CQ current is
\bea
j_{I}^\mu= j_{1}\gamma^\mu +j_{2} \frac{i\sigma^{\mu\nu}q_\nu}{2M} \, , \label{CQc}
\eea
where $j_1$ and $j_2$ both depend on isospin ($\tau_3=\pm$) and $Q^2$ through the CQ form factors 
\bea
j_1&=&f_{1+}+\tau_3 f_{1-} \,  ;  \qquad f_{1\pm}= {\textstyle\frac{1}{3}}f_{1u}\mp{\textstyle\frac{1}{6}}  f_{1d}
\nonumber\\
j_2&=&f_{2+}  +\tau_3 f_{2-} \, ;  \qquad f_{2\pm}= {\textstyle \frac{1}{2} } \left(\mu_uf_{2u}\pm\mu_df_{2d}\right)\, .\qquad
\eea
Here $\mu_u$ and $\mu_d$ are the anomalous moments of the $u$ and $d$ quarks, adjusted to fit the nucleon anomalous moments.% {\it exactly\/}.
 
Three principals guide the choice of CQ form factors. First, QCD requires that the CQ  should become elementary point-like quarks as $Q^2\to\infty$, and this requires  $f_{2\pm}(Q^2)\to 0$ and $f_{1\pm}(Q^2)\to\lambda_\pm$ as $Q^2\to\infty$.  To assure that the  charge assignments remain in the ratio of 2/3 to $-1/3$ at both large $Q^2$ (when the quarks are point-like and the charge assignment is given by QCD) and small $Q^2$ (where the charge assignments are given by quark model phenomenology) we require $\lambda_+=\lambda_-\equiv\lambda$.  [The charges could then be changed to $2/3$ and $-1/3$ at $Q^2\to\infty$ by absorbing $\lambda$ into $N_0^2$.]  Second, guided by vector dominance, we expect the individual isoscalar $f_{i+}$ and isovector $f_{i-}$ form factors to have a monopole contribution
\bea
G(\Lambda^2)= \frac{\Lambda^2}{\Lambda^2+Q_0^2} 
\eea
with 4 scale parameters, $\Lambda_{i\pm}^2$, independently adjusted to fit the data.  We find the effective vector dominance masses to be strongly channel dependent (see Table \ref{tab1}).  Third, we allow for direct coupling of the photon to pions inside the nucleon by adding a purely  isovector dipole charge contribution to $f_{1-}$
\bea
G_\pi=\frac{\Lambda_\pi^4}{\left(\Lambda_\pi^2+Q_0^2\right)^2}\, .
\eea
Assembling these pieces gives the following  
\bea
&&f_{1+}(Q^2)=e_+[(1-\lambda) G(\Lambda_{1+})+\lambda]\cr
&& f_{1-}(Q^2)=e_-[(1-\lambda-\lambda_\pi)G(\Lambda_{1-})+\lambda_\pi G_\pi+\lambda]\, ,\qquad\cr
&&f_{2\pm}(Q^2)=\mu_\pm G(\Lambda_{2\pm})  
\eea
where $e_+=\frac{1}{6}$, $e_-=\frac{1}{2}$, $\mu_\pm=\frac{1}{2}(\mu_u\pm\mu_d)$.
%, and the elementary functions are all normalized to unity at $Q^2=0$

To complete the calculation the second term in (\ref{eq8}) must be reduced.  The isospin algebra yields
\bea
 j_{(i+2)}&\equiv&{\textstyle \frac{1}{3}}\tau_j j_{i}Ê\tau_j= f_{i+} -{\textstyle \frac{1}{3}}\tau_3 f_{i-}  \, ,
\eea
where $i=\{1,2\}$.
Averaging over the asimuthal angle and using the Dirac equation allows the extraction of the nucleon form factors $F_1$ and $F_2$
%\begin{widetext}
%
\bea
F_1(Q^2)&=&\frac{3}{2}\int \frac{m_s^2\,d^3
\mathbb{\kappa}}{(2\pi)^3 2E_s(\kappa)} \Big\{
\psi^+_0\psi^-_0 \;  j_1 
\cr&&\qquad
+\frac{1}{3}\psi^+_1\psi^-_1  \left( j_3R_1 - j_4{\textstyle\frac{1}{4}}Q_0^2R_2\right)\Big\}\cr
F_2(Q^2)&=&\frac{3}{2}\int \frac{m_s^2\,d^3
\mathbb{\kappa}}{(2\pi)^3 2E_s(\kappa)} \Big\{
\psi^+_0\psi^-_0 \;  j_2 
\cr&&\qquad
- \frac{1}{3}\psi^+_1\psi^-_1  \left( j_3 R_2+ j_4 R_3\right)\Big\} \, ,
\eea
%
%\end{widetext}
where $R_1$, $R_2$, and $R_3$ are
\bea
\left(1+{\textstyle\frac{1}{4}}Q_0^2\right)R_1&=&3+2\kappa^2+{\textstyle\frac{1}{4}}Q_0^2[1-\kappa^2(1-z^2)] \cr
\left(1+{\textstyle\frac{1}{4}}Q_0^2\right)R_2&=& 2 +\kappa^2(3-z^2)\cr
R_3&=& 1+\kappa^2(1+z^2) -{\textstyle\frac{1}{4}}Q_0^2R_2\, .
\eea
At $Q^2=0$ the relativistic factor (\ref{BA}) cancels $R_1$, giving
\bea
F_1(0)=\frac{3}{2}\left(j_1+j_3\right)I_1= \frac{1}{2}[1+\tau_3]\,I_1 \label{norm}
\eea
with $D_\pm(0)=D(0)$ and
\bea
I_1=\int \frac{d^3
\mathbb{\kappa}}{(2\pi)^3 2E_s(\kappa)} \left(\frac{N_0^2}{D^2(0)}\right)\, .
\eea
The condition $I_1=1$ now insures that the correct proton and neutron charges emerge automatically.  In a full theory this feature would emerge without additional assumptions \cite{Gross:2003qi,Adam:1997rb} so we do not regard $N_0$ as an adjustable parameter.

%%%%%%%%%%%%%%%%%%%%%%%%
\begin{figure}
%\vspace*{-1in}
%\begin{center}
\centerline{
\mbox{
\includegraphics[width=3in]{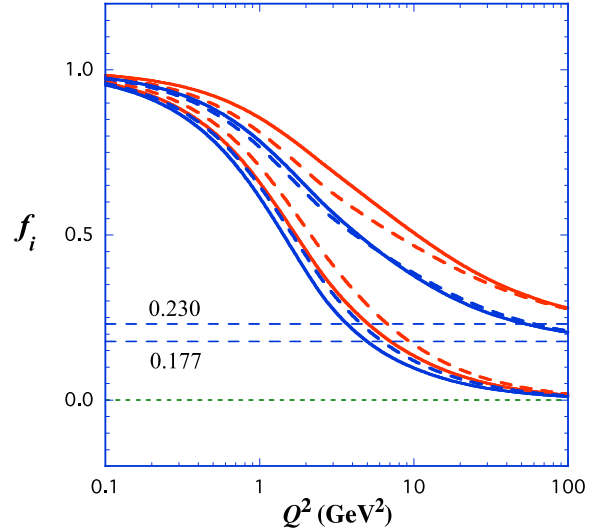}
}
}
%\end{center}
%\vskip -1.0in
%\vskip -0.8in
\caption{\footnotesize\baselineskip=10pt (Color on line) The quark form factors.  Solid lines are $u$ and dashed lines are  $d$ form factors; upper 4 lines are $f_1$ (Model I larger than II) and  lower 4 lines $f_2$ (Model I again larger than II). Asymptotic values of the $f_1$ form factors are shown. }
\label{fig3}
\end{figure} 
%%%%%%%%%%%%%%%%%%%%%%%%
%%%%%%%%%%%%%%%%%%%%%%%%

As $Q^2\to\infty$, the form factors $f_{1\pm}$ approach $\lambda e_\pm$ and the form factors $f_{2\pm}$ approach zero as $\Lambda_{2\pm}^2/Q_0^2$.  Then
\bea
&& \lim_{Q^2\to\infty}F_1=\frac{3\lambda
N_0^2}{2Q_0^4}  
[e_+ + \tau_3 e_-]\, I_2 \cr &&\cr
&&\lim_{Q^2\to\infty}F_2 = \frac{3N_0^2}{2Q_0^6} [m_+ +\tau_3 m_- ]\,I_2 \label{eq26}
\eea
where $m_\pm\equiv\mu_\pm\Lambda_{2\pm}^2$ and $I_2$ is
\bea
I_2=\int \frac{d^3
\mathbb{\kappa}}{(2\pi)^3 2E_s(\kappa)} 
\frac{1}{[1+\kappa^2(1-z^2)]^2}=0.0179%334
\, . \quad
\eea

%%%%%%%%%%%%%%%%%%%%%%
\begin{table}
\begin{center}
\caption{Ten adjustable parameters (all dimensionless).  The 11th adjustable parameter, $\lambda$, is 0.230 (I) or  0.177 (II).}
\begin{minipage}{3.5in}
\begin{tabular}{cccccc}Model &
$\;\beta_1$, $\beta_2\;$ & $\;\;\mu_u$, $\mu_d\;\;$ & $\;\;\Lambda^2_{1+}$, $\Lambda^2_{1-}\;\;$ & $\;\Lambda^2_{2+}$, $\Lambda^2_{2-}\;$ &$\;\lambda_\pi$,  $\Lambda_\pi^2$\cr
\hline%\cr
I & $0.057$&$\phantom{-}$1.125&$\quad$7.69$\quad$ &$\quad$0.362$\quad$& 0.245 \cr
&$1.982$&$-0.837$&10.67&1.82 &2.15\cr
%\hline
II & $0.069$ & $\phantom{-}$1.126&3.23&0.386&0.244 \cr
& $1.543$  &$-0.825$&5.81&1.23&$0.742$%\cr
%\hline
\end{tabular}
\label{tab1} 
\end{minipage}
\end{center}
\vspace*{-0.3in}
\end{table}
%%%%%%%%%%%%%%%%%%%%%%%%%%%%%%%%%%%%%%%%%%%%%%%%%%%

%%%%%%%%%%%%%%%%%%%%%%
\begin{table}
\begin{center}
\vspace*{-0.1in}
\caption{Properties of the two fits. For comparison, the experimental values are $r_p^2=0.780(25)$ and  $r^2_n = -0.113(7)$.}
\begin{minipage}{3.5in}
\begin{tabular}{ccccc}Model$\quad$ & $\chi^2$/datum & 
$r_p^2 ({\rm fm}^2)$  & $r_n^2 ({\rm fm}^2)$  & $N_0^2$ \cr
\hline%\cr
I & 1.18&0.800&$-0.065$ & 36.39 \cr
II &$\quad$1.94$\quad$ &$\quad$0.771$\quad$&$\quad-0.106\quad$ &$\quad 60.87 \quad$
\end{tabular}
\label{tab2} 
\end{minipage}
\end{center}
\vspace*{-0.4in}
\end{table}
%%%%%%%%%%%%%%%%%%%%%%%%%%%%%%%%%%%%%%%%%%%%%%%%%%%

\begin{figure}[t]
%\vspace*{-1in}
%\vspace*{-0.8in}
%\begin{center}
\centerline{
\mbox{
\includegraphics[width=3.9in]{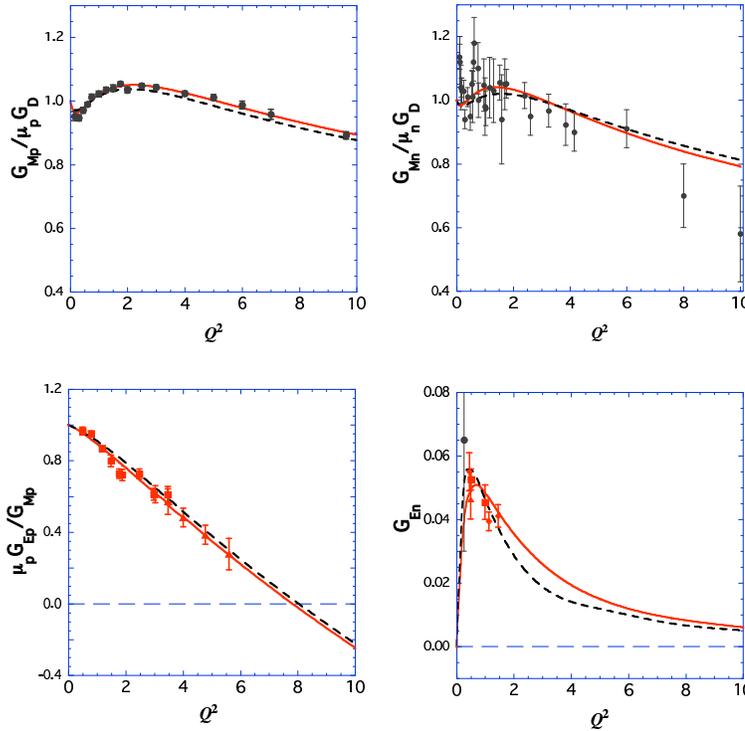}
}
}
%\end{center}
%\vskip -2.0in
%\vskip -1.6in
\caption{\footnotesize\baselineskip=10pt (Color on line) Data for the nucleon form factors compared with Model I (solid lines) and Model II (dashed lines).  The $G_{Mp}$ and $G_{Mn}$ data were used by Bosted in his global fits to the form factors \cite{Bosted}.  The $G_{Ep}$ data are from JLab Hall A by Jones, \etal \cite{Jones} (squares) and  Gayou, \etal \cite{Gayou} (triangles).  The $G_{En}$ data are from Bates \cite{Bates} (circle),  and from JLab Hall C by Zhu, et.al. \cite{Zhu} (triangle), Warren, \etal \cite{Warren} (squares), and Madey, \etal \cite{Madey}(diamonds).  }
\vspace*{-0.2in}
\label{fig1}
\end{figure}

We present two fits to the nucleon form factors.  The parameters of each fit are shown in Table \ref{tab1}, with properties displayed in Table \ref{tab2},  quark  form factors shown in Fig.~\ref{fig3}, and nucleon form factors shown in Fig.~\ref{fig1}.  (We have not used data for $G_{Ep}$ obtained from Rosenbluth separations because it is now believed that such data are contaminated by two photon exchange contributions \cite{Chen:2004tw}.)  Model I was obtained by minimizing $\chi^2$, and gives excellent agreement (see Table \ref{tab2}).    Model II starts from Model I and further adjusts the parameters to give an excellent fit to the charge radii.

The two magnetic form factors $G_{Mp}$ and $G_{Mn}$ have a very similar shape, which seems to disagree with the high $Q^2$ trend for $G_{Mn}$.  However, this ``trend'' is defined by only two data points with large errors.  These models predict the shape that new, more accurate measurements of $G_{Mn}$ (being analyzed \cite{brooks}) should follow.  The fits to the electric form factors are both excellent, and predict that  $G_{Ep}$ will go through zero at $Q^2\sim 8$ GeV$^2$.  The quality of our fits is comparable to the recent vector dominance model of Ref.~\cite{Bijker:2004yu}. 

%%%%%%%%%%%%%%%%%%%%%%%%
%%%%%%%%%%%%%%%%%%%%%%%%
%%%%%%%%%%%%%%%%%%%%%%%%

%%%%%%%%%%%%%%%%%%%%%%%%
\begin{figure}[b]
%\vspace*{-1in}
%\vspace*{-0.75in}
%\begin{center}
\centerline{
\mbox{
\includegraphics[width=3.7in]{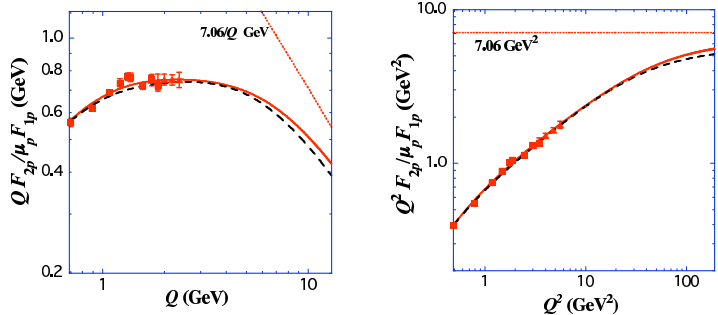}
}
}
%\end{center}
%\vskip -4.5in
%\vspace*{-3.3in}
\caption{\footnotesize\baselineskip=10pt (Color on line) Double log plots showing the data for $QF_{2p}/(\mu_pF_{1p})$ and $Q^2F_{2p}/(\mu_pF_{1p})$ versus $Q$ and $Q^2$ respectively (labeled as in Fig.~\ref{fig1}).  The dotted lines are the asymptotic prediction for Model I, $Q^2F_{2p}/(\mu_pF_{1p})=7.06$, obtained from Eq.~(\ref{eq26}).  }
\label{fig2}
\end{figure} 
%%%%%%%%%%%%%%%%%%%%%%%%
%%%%%%%%%%%%%%%%%%%%%%%%

The observation that $QF_{2p}/F_{1p}$ is nearly flat for the highest points measured has lead to speculation that  high energy $ep$ elastic scattering violates the conservation of helicity \cite{Ralston:2003mt}, and that the proton may have an exotic electromagnetic structure.   Fig.~\ref{fig2} shows that our Models predict that the flatness currently observed is only temporary, and that $QF_{2p}/F_{1p}$ does indeed approach zero as $1/Q$, as expected.  The asymptotic behavior is delayed until $Q^2>100$ GeV$^2$.  This conclusion is largely independent of the running of the QCD coupling constant $\alpha_s$,  which we have not discussed.

We conclude that the new form factor data do not provide evidence that the nucleon is anything other than a spherically symmetric state of three relativistic constituent quarks \cite{Li:1975am}.  We also found that the data can {\it not\/} be fit  {\it unless\/} it is assumed that the CQ have an electromagnetic structure.  The precision of the data allow this structure to be determined empirically (subject to uncertainties as shown in Fig.~\ref{fig3}).  This fit requires a pion ``cloud'' (the $G_\pi$ term) and leads to the conclusion that the onset of perturbative QCD is delayed until large $Q^2$, as expected.

It is a pleasure to acknowledge helpful conversations with Peter Bosted and Edward Brash, both of whom shared their data files with us, and to thank  J.M. Laget and C.~Weiss for their careful reading of the manuscript.   This work was
supported in part by the US Department of  Energy  under grant
No.~DE-FG02-97ER41032. The Southeastern Universities Research Association
(SURA) operates the Thomas Jefferson National Accelerator Facility under DOE
contract DE-AC05-84ER40150.

\end{document}